\documentclass[12pt,preprint]{aastex}

\newcommand{\EUVE}{{\it EUVE}}
\newcommand{\Boo}{44{\it $\iota$\/}~Boo}

\newcommand{\ROSAT}{{\it ROSAT}}

\newcommand{\Chandra}{{\it Chandra}}

\begin{document}


\title{X-ray Doppler Imaging of \Boo\ with \Chandra}

\author{N. S. Brickhouse, A. K. Dupree, and P. R. Young}
\affil{Harvard-Smithsonian Center for Astrophysics, 60 Garden
Street, Cambridge, MA  02138}
\email{nbrickhouse@cfa.harvard.edu, adupree@cfa.harvard.edu, pyoung@cfa.harvard.edu}

\doublespace

\begin{abstract}

\Chandra\ High-Energy Transmission Grating observations of the bright
eclipsing contact binary \Boo\ show X-ray line profiles which are
Doppler-shifted by orbital motions. The X-ray emission spectrum
contains a multitude of lines superimposed on a weak continuum, with
strong lines of \ion{O}{8}, \ion{Ne}{10}, \ion{Fe}{17}, and
\ion{Mg}{12}.  The profiles of these lines from the total observed
spectrum show Doppler-broadened widths of $\sim 550$ km s$^{-1}$. Line
centroids vary with orbital phase, indicating velocity changes of $>
180$ km s$^{-1}$.  The first-order light curve shows significant variability,
but no clear evidence for either primary or secondary eclipses. Flares
are observed for all spectral ranges; additionally, the light curve
constructed near the peak of the emission measure distribution ($T_e =
5$ to $8 \times 10^6$ K) shows quiescent variability as well as
flares.  The phase-dependences of line profiles and light curves
together imply that at least half of the emission is localized at high
latitude.  A simple model with two regions on the primary star at
relatively high latitude reproduces the observed line profile shifts
and quiescent light curve.  These first clear X-ray Doppler shifts of
stellar coronal material illustrate the power of \Chandra.

\end{abstract}


\section{Introduction}

We report the first ``X-ray Doppler imaging'' of a late-type binary
system, made possible by the high spectral resolution and stable
wavelength scale of the \Chandra\ High-Energy Transmission Grating
(HETG).  Phase-dependent line profile measurements of the contact
binary \Boo, together with temperature-selected light curves, provide
powerful new diagnostic capability.

Optical Doppler imaging of rapidly rotating RS CVn systems has
demonstrated coronal activity patterns distinctly different from those
of the Sun, with relatively stable polar active regions co-existing
with more transient equatorial active regions (e.g. Vogt et al. 1999;
Donati \& Collier Cameron 1997). While optical Doppler imaging has
been applied to RS CVn systems for more than a decade, only recently
have Doppler images of a contact binary been reported, for the highly
variable VW~Cep, showing the presence of large polar spots, as well as
low latitude features (Hendry \& Mochnacki 2000).

In X-rays, the spectral resolution has until now been inadequate for
line profile analysis of late-type stars. Eclipse mapping from light
curve analysis has been used to infer the sizes and locations of
flares (e.g. on Algol, Schmitt \& Favata 1999; and on VW Cep, Choi \&
Dotani 1998); however, it is generally difficult to disentangle
variability due to flaring from the modulation produced by either
rotation or eclipses.  The \EUVE\ Deep Survey light curve for \Boo\
over 19.6 epochs shows no eclipses, but reveals quiescent modulation
at the 50\% level, associated with the rotation of the primary star;
furthermore, EUV density diagnostics from the \EUVE\ spectrometers
suggest high density, implying localized coronal regions with small
scale size $l \sim .004 R_{\star}$, which must be located at high
latitude (Brickhouse \& Dupree 1998; hereafter, BD98). Determining the
sizes and locations of coronal structures on contact binaries is an
important step toward understanding the rotation/activity correlation,
given that contact binaries are anomalously underluminous in X-rays,
presenting normalized X-ray fluxes ($L_x/L_{bol}$)  4 -- 5 times weaker than those of the fastest
rotating single stars (St{\c{e}}pien, Schmitt, \& Voges 2001).

\section{Observations}

\Boo\ (HD~133640) is a triple star system containing a contact binary
of the W~UMa type, located a distance of 12.76 pc (ESA 1998). X-ray
emission, first observed by Cruddace \& Dupree (1984), is expected to
originate with the contact binary components B and C of the
system. These two stars, of type G0 Vn (Gray, Napier, \& Winkler 2001),
orbit with a period of 6.4 hours and an
inclination of 72.8$^\circ$ (Hill, Fisher, \& Holmgren 1989).

Time-tagged HETG spectra of \Boo, using the Advanced CCD Imaging
Spectrometer-S (ACIS-S) detector, were obtained by \Chandra\ on 2000
April 25 (ObsID 14). The system was observed for 59142 s, continuously
covering 2.56 epochs of the orbit. The event lists were
obtained from the Chandra X-ray Observatory Center (CXC) public
archive as Level 1 files, and were then processed, using CXC software CIAO V2.1 and
calibration database CALDB V5.0. High-Energy Grating (HEG) and
Medium-Energy Grating (MEG) spectra were extracted as described by
Canizares et al. (2001), and source-specific effective area curves
were generated.  Positive and negative first-order spectra were
co-added.

\section{Data Analysis}

Here we link phase-dependent changes in the line profiles to the
phase-binned behavior of the light curve, by applying the
techniques of both eclipse mapping and X-ray Doppler imaging.
Emission lines from the available Fe ionization stages provide broad
coverage of electron temperature $T_e$ and allow construction of a
continuous emission measure distribution (EMD) model. The model EMD
allows us to determine the $T_e$ range
contributing to the emission lines of interest for profile analysis,
as well as to predict which other spectral features arise from the same
$T_e$ range. Photons from all of these features are then selected from the
event list to construct a new light curve characteristic of the
line-emitting $T_e$.

\subsection{Total Spectrum and Emission Measure Distribution}

The total first-order spectrum of \Boo\ shows strong lines of H-like
and He-like O, Ne, Mg, and Si, along with Fe L-shell lines from
several ionization states. A weak continuum is also apparent. Table~1
gives the measured line fluxes and linewidths for the four strongest
emission lines (\ion{O}{8}, \ion{Ne}{10},
and \ion{Mg}{12} Ly$_{\alpha}$ and \ion{Fe}{17} $2p^53d~^1P_1 - 2p^6~^1S_0$),
as determined from Gaussian fits to the
continuum-subtracted line profiles using the {\it Sherpa} package in
CIAO. Line profiles are wider than the instrumental FWHM (240 and 460
km s$^{-1}$ at 15 \AA, for HEG and MEG, respectively), with widths of $\sim$
550 km s$^{-1}$ corresponding to the spread of velocities in the system.

Strong, isolated Fe lines are used to construct the EMD: \ion{Fe}{17} $\lambda
\lambda$16.780; 17.051; 17.096, \ion{Fe}{18}I $\lambda \lambda$14.208;
15.625, \ion{Fe}{19} $\lambda$14.664, \ion{Fe}{22} $\lambda$11.770,
and \ion{Fe}{23} $\lambda$11.736. The
method is described by BD98 and we use the spectral models from the Astrophysical
Plasma Emission Code (APEC; Smith et al. 2001).  The EMD is characterized by a narrow
peak at $8 \times 10^6$ K, similar to that found by BD98, but with the
peak emission measure about 20\% lower and the emission measure for
$T_e > 10^7$ K about 50\% higher.

Although the four strongest lines potentially represent a broad range
of $T_e$, as characterized by their temperature of maximum emissivity
$T_{max}$ (see Table~1), in the \Boo\ spectrum they are all formed
within a much narrower range. The EMD bump over the $T_e$ range, $T_L
= 5$ to $8 \times 10^6$ K, produces more than half of the flux for
each of these lines (65, 83, 92, and 68\%, respectively, for
\ion{O}{8}, \ion{Ne}{10}, \ion{Fe}{17}, and \ion{Mg}{12}).  Models
show that the spectral region between 14 and 20 \AA\ is primarily
composed of line emission produced at $T_L$, and thus photons from
this spectral region are used to construct the $T_L$ light curve.

\subsection{Light Curves}

Figure~1 shows the optical light curve of Gherega et al. (1994), which
has subsequently been confirmed by eclipse timing measurements
(Pribulla, Chochol, \& Parimucha 1999; Albayrak \& Gurol 2001).
Primary eclipse (phase 0.00) corresponds to the eclipse of the
secondary by the primary. Also shown are the \Chandra\ total
first-order light curve and the $T_L$ light curve, limited to photons
falling between 14 and 20 \AA.  The total light curve shows
significant variability, but no clear evidence for either primary or
secondary eclipses. A light curve constructed for high energy photons
($\lambda < 8$ \AA; not shown) tracks the features of the total light
curve, suggesting that the largest variations in total count rate
are produced at the highest temperatures in the corona. The $T_L$
light curve shows a different pattern of variability.

Four flares are identified in the total light curve, with the largest
flare producing a peak count rate 3 times higher than the total count
rate minimum. The $T_L$ light curve shows three of the flares, with
the largest flare producing only twice the count rate at peak compared
to the $T_L$ count rate minimum.

Excluding the times of obvious flares, the $T_L$ light curve is
notable in several respects which may reflect its ``quiescent''
behavior: (1) The $T_L$ light curve is variable at the $\sim$20\%
level, consistent with the X-ray variability found with \ROSAT\
(McGale, Pye, \& Hodgkin 1996). (2) Narrow dips occur immediately
following each of the three primary eclipses, with remarkably similar
shapes on each occurrence (Fig.~1). These dips indicate that emission
is being absorbed, eclipsed, or rotated out of view. (3) A somewhat
broader dip is observed near one of the two secondary eclipses that
occur during the \Chandra\ pointing. This dip may also indicate
occultation or a region rotating out of view; however, such an
interpretation is not well supported because the other secondary eclipse
coincides with a flare, observed in the total light curve, but not in
the $T_L$ light curve. Perhaps a dip effectively masks the $T_L$
flare. (4) Excluding flares and dips, the $T_L$ light curve is
consistent with the \EUVE\ finding (BD98) of sinusoidal variation with
only one maximum, rather than two maxima as found in the optical light
curve. Periodicity at or near the optical period, as found in the 19.6
epoch long \EUVE\ observation, is not confirmed.

\subsection{Phase-binned Line Profiles}

Figure~2 shows the \ion{Ne}{10} and \ion{Fe}{17} line profiles summed
over four equal phase intervals centered at Phase 0.00, 0.25, 0.50,
and 0.75. Also shown are instrumental line profiles, centered at the
laboratory wavelengths.  While the MEG profiles have greater
signal-to-noise, the factor of two greater resolving power of HEG
allows much better isolation of different velocity components.  The
\ion{O}{8} and \ion{Mg}{12} profiles (not shown) are similar, but with
lower spectral resolution (HEG does not cover the \ion{O}{8} line) or
signal-to-noise ratio.

The rest wavelengths of these lines are well known. The H-like
Ly$_{\alpha}$ wavelengths are taken as the weighted
average of the doublet, assuming they are optically thin and
collisionally excited; the \ion{Fe}{17} wavelength is 15.014 $\pm$
.001 \AA\ (Brown et al. 1998). The absolute wavelength scale of the
HETG is currently only accurate to about 100 km s$^{-1}$; however, the scale
appears to be very stable, such that {\it relative} wavelength
measurements are secure to better than 100 parts per million, or about
30 km s$^{-1}$ (CXC Proposers' Observatory Guide 2001).  The reality of
individual features in the line profiles is more difficult to assess,
but we expect that significant deviations from a Gaussian profile,
occurring over several bins, reflect real features. The profiles
contain multiple components representing a range of
velocities. Roughly one third of the emission is always present within
the instrumental line profile, and cannot be resolved.  Excluding data
during the flare times (Fig.~1) does not affect these components.

Median wavelengths are determined from MEG line profiles as the
interpolated center of the count distribution, with Poisson errors
assumed. Continuum and weaker background are not subtracted, as they
appear to contribute at most 20 counts to any of the total integrated
MEG lines.  Figure~3 shows the shift of the median wavelengths for
each line, for the same four orbital phase bins as in
Figure~2. Centroids determined from separate MEG and HEG fits to
Gaussian line profiles are consistent with these medians, but give
larger errors where the profiles appear distinctly non-Gaussian.  The
four ions studied show similar patterns of velocity shifts as a
function of orbital phase. Relative velocity changes with phase exceed
the instrumental uncertainties, and suggest net velocity changes of
180 km s$^{-1}$ over the orbit. \ion{Ne}{10} has sufficient counts to verify
that the patterns of line shifts repeat from one epoch to the next.

\section{Modeling the Coronal Activity}

We present a simple 2-component model of the variability due to
quiescent coronal structures. The lack of eclipse signatures (two
maxima and two minima per orbit) severely limits the amount of
emitting material that can be evenly distributed near the surfaces of
both stars, since the equatorial regions would then be eclipsed.
Diffuse extended emission is also strictly limited, since the
integrated broad line profiles show distinct velocity components which
change with phase. The $T_L$ light curve shows about 20\% quiescent
modulation, implying that most of the material is always visible,
i.e. located at high latitude. Flares are not correlated with phase,
consistent with the active regions being visible most of the time.

Our model attempts to match the observed features: (a) phase-dependent
line shifts; (b) broad sinusoidal $T_L$ light curve modulation found
from \EUVE\ (BD98) and confirmed with the present observation; and (c)
narrow dips in the $T_L$ light curve. Two regions contribute roughly
equally to the total emission in order to account for similar drops in
the count rate during the narrow dips and broad minima.  Since
line centroids show velocity shifts consistent with the motion of the
primary star, we place both active regions on the primary star. They
are assumed to lie on the star's surface. The distinctive dip in the
light curve just after phase 0.0 can only be explained as a small
emitting region R1 on the inner face of the primary, which briefly
rotates out of view.  R1 must extend below a latitude of 72.8$^\circ$
in order to be occulted, but not extend below $\approx$ 70$^\circ$ so
that the dip remains narrow. (Of course, if the material is extended
above the stellar surface, the latitude constraint is weakened.) For
the same reason R1 cannot extend more than $\approx 10^\circ$ in
longitude.

To reproduce the broader modulation requires a larger region R2 on the
outer face of the primary in order to yield a light curve maximum at
phase $\approx 0.2$. R2 must also be at high latitude in order to
produce the weak $\sim 20$\% modulation.  Figure~4 shows that the
light curve produced from a particular choice of the regions R1 and R2
mimics the key features of the Chandra $T_L$ light curve. Simulations
of the system with different active region locations and sizes,
including radially extended regions, are underway to study the effects
on light curves and line profiles.

\section{Conclusions}

The \Chandra\ observations of \Boo\ provide the first clear evidence
from line profile velocity measurements for localized activity at high
latitude on the primary star. Our interpretation is supported by the
phase-dependent light curves for emission produced at the temperature
of the EMD bump. Location of activity on the primary star has been
considered in the past to explain peculiarities and changes in the
photometric behavior of W UMa binaries, although a strong theoretical
basis for such a preference is lacking (Rucinski 1985). Coronal
enhancements that we find associated with the primary provide added
support for the dominance of the primary in magnetic activity as other
studies have indicated (BD98; Choi \& Dotani 1998; Barden 1985). These results demonstrate the power of \Chandra\ for
understanding coronal structure, using eclipse mapping and X-ray
Doppler imaging.

\acknowledgements

This work was supported in part by NASA NAG5-3559 and the Chandra
X-Ray Observatory Center NASA NAS8-39073 from NASA to the Smithsonian
Astrophysical Observatory and by Johns Hopkins Contract \#2480-60016.

\clearpage

\begin{deluxetable}{lrrrcccc}
\small
\tablecolumns{10}
\tablewidth{0pc}
\tablecaption{Strong Line Measurements}
\tablehead{
\colhead{Ion} & 
\colhead{$\lambda$} & 
\colhead{MEG} &
\colhead{HEG} &
\colhead{Flux} &
\colhead{MEG FWHM} &
\colhead{HEG FWHM} & 
\colhead{$T_{max}$}  \\
\colhead{} &
\colhead{(\AA)} &
\colhead{(Cnts)} &
\colhead{(Cnts)} &
\colhead{(ph cm$^{-2}$ s$^{-1})$} &
\colhead{(m\AA)} &
\colhead{(m\AA)} &
\colhead{(K)}  }
\startdata
\ion{O}{8}                     & 18.969 & 721  & \nodata  & $1.21 \pm 0.05 \times 10^{-3}$ & 34.2 $\pm$ 1.4 & \nodata         & $3 \times 10^6$  \\
\ion{Ne}{10}\tablenotemark{a}      & 12.134 & 1544 & 434      & $5.18 \pm 0.14 \times 10^{-4}$ & 29.4 $\pm$ 0.1 & 23.4 $\pm$ 1.4  & $5 \times 10^6$  \\
\ion{fe}{17}                    & 15.014 & 401  & 121      & $3.87 \pm 0.23 \times 10^{-4}$ & 33.2 $\pm$ 2.3 & 25.4 $\pm$ 3.6  & $5 \times 10^6$  \\ 
\ion{Mg}{12}                     &  8.421 & 347  & 163      & $5.94 \pm 0.38 \times 10^{-5}$ & 23.7 $\pm$ 1.7 & 17.2 $\pm$ 2.0  & $8 \times 10^6$  

\tablenotetext{a}{The contribution of Fe XVII $\lambda$12.124 is
estimated to be only  $\sim 7$\%, based on ratios with \ion{Fe}{17} $\lambda$17.051.}
\enddata
\normalsize
\end{deluxetable}

\clearpage
\begin{figure}
\plottwo{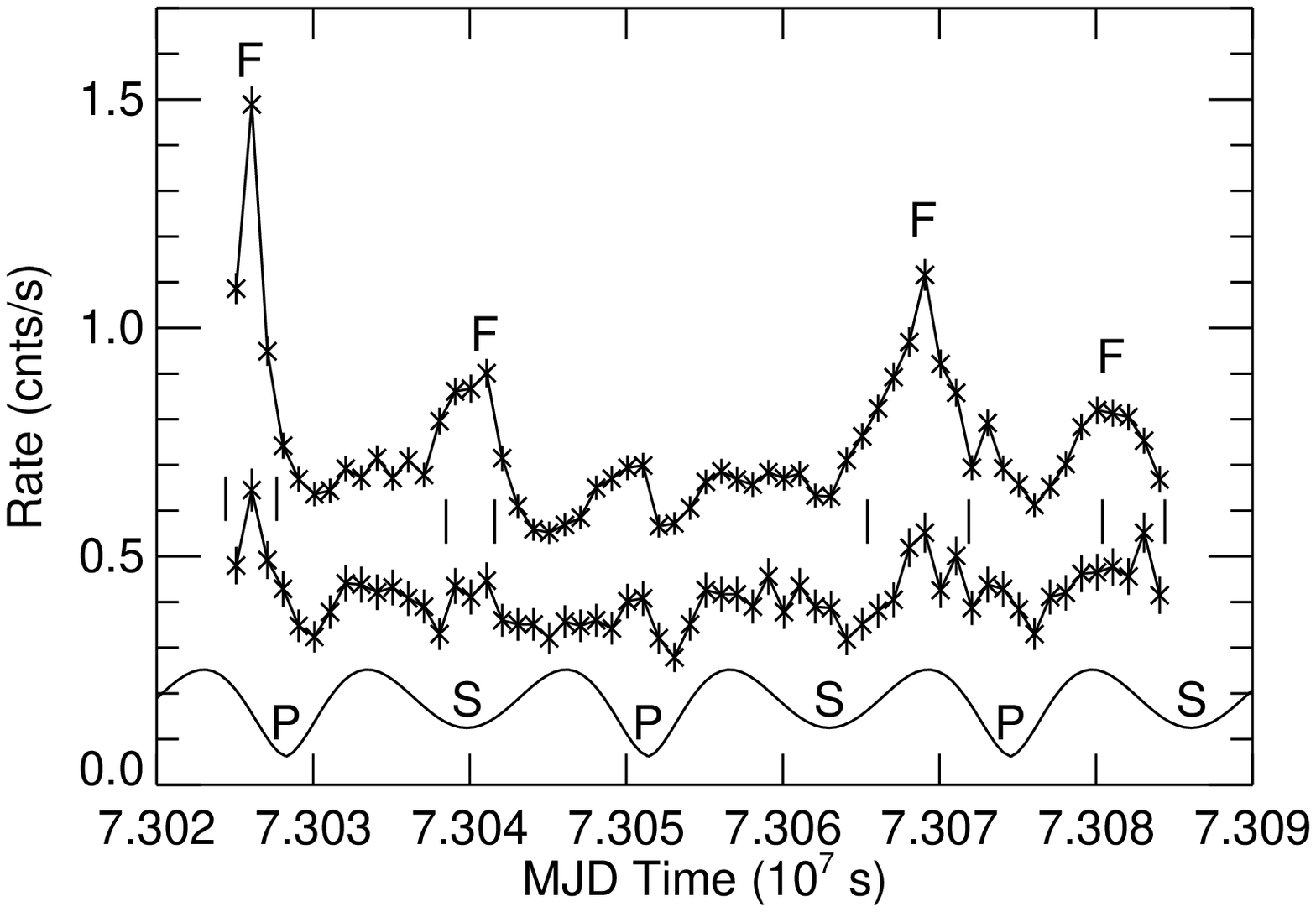}{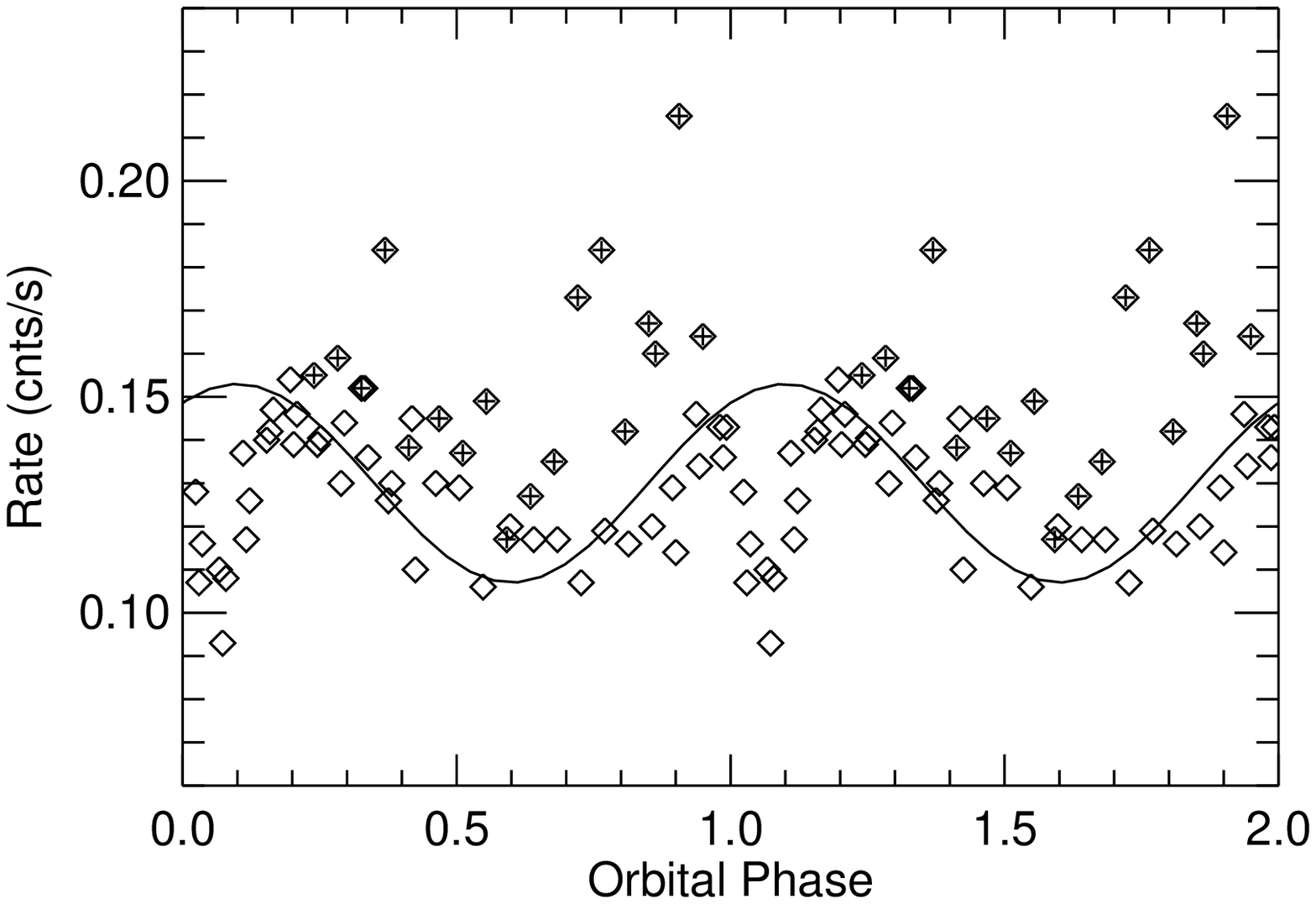}
\caption{{\it Left}: Count rate vs MJD (s) for 1000 s time
intervals. The uppermost curve is the total first-order light
curve. The middle curve represents the low temperature $T_L$ ($14$
\AA\ $< \lambda < 20 $ \AA) light curve (count rate $\times 3$). The
lowest curve is the scaled optical light curve, with the primary (P)
and secondary (S) minima marked. Four flares (F) are identified, with
vertical bars denoting their extent. {\it Right}: Count rate vs
orbital phase for the same $T_L$ light curve shown above. Data
($\diamond$) are phase-folded over two epochs. Plus symbols ($+$) are
overplotted for the flare times marked with vertical bars on the
figure above ({\it left}). A sine curve is overlaid on the light curve
for illustration.
}
\end{figure}

\begin{figure}
\plottwo{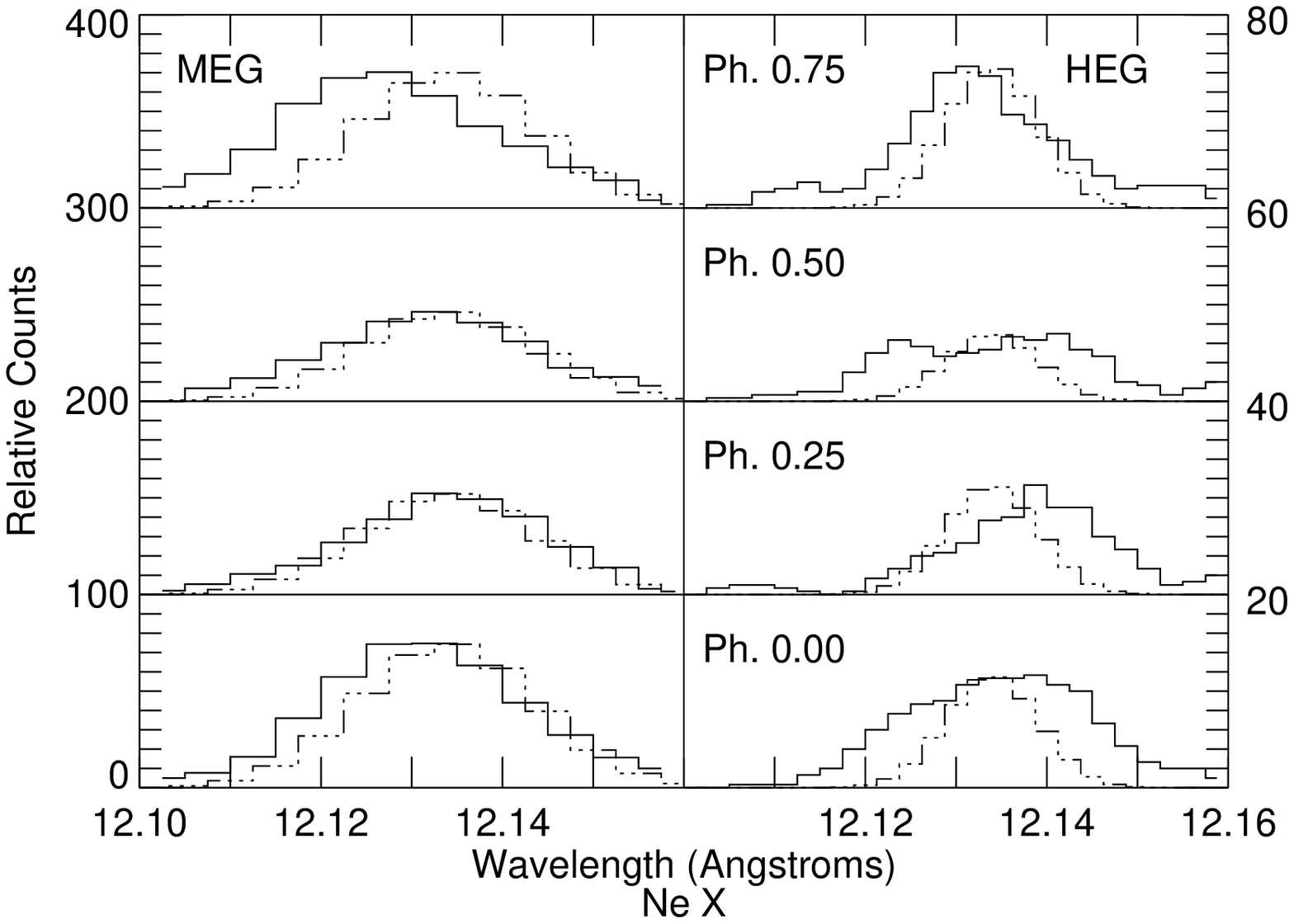}{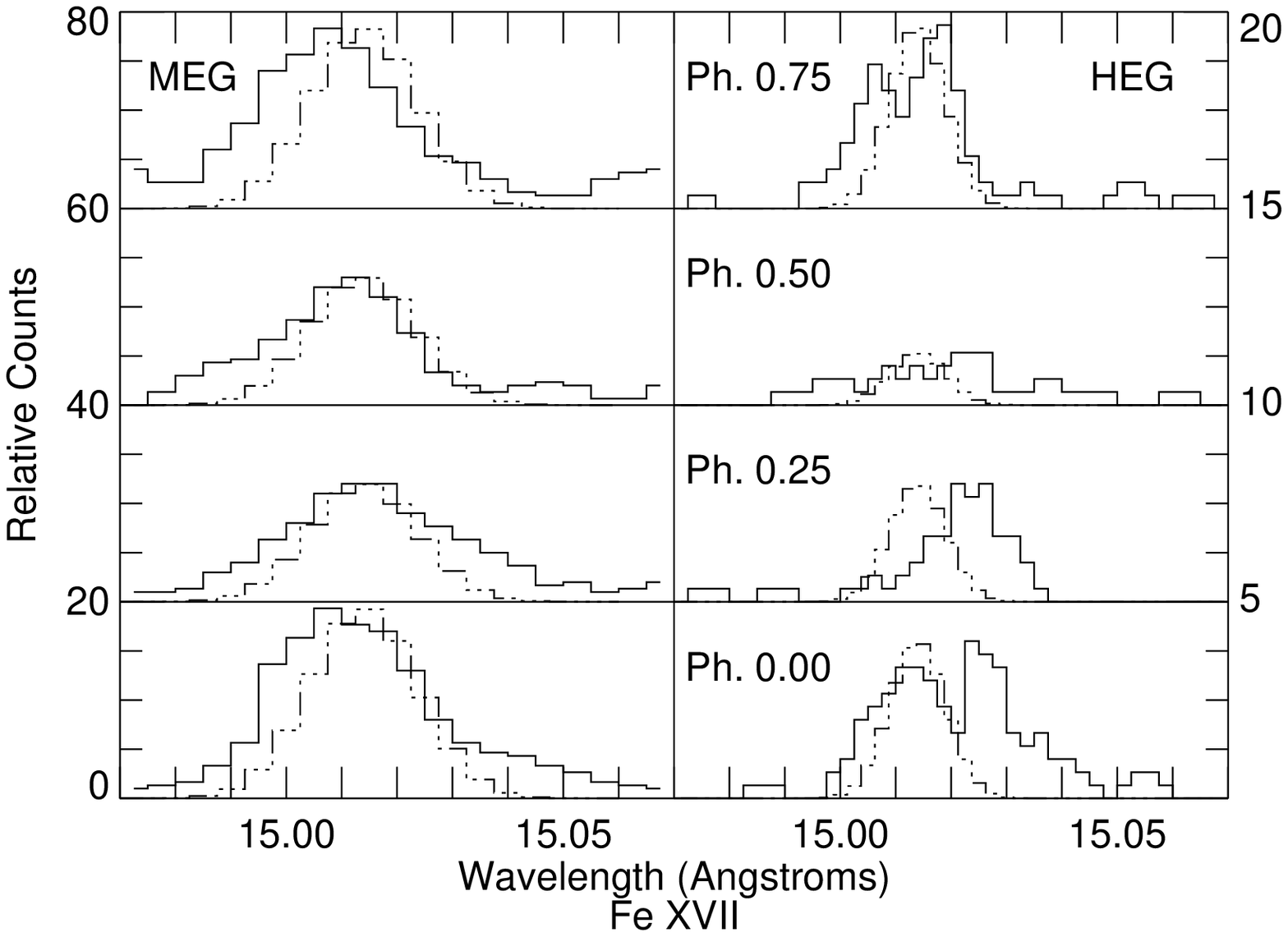}
\caption{\ion{Ne}{10} ({\it left}) and \ion{Fe}{17} ({\it right}) MEG
and HEG line profiles (solid) for each phase. Binsizes are 0.005 and
0.0025 \AA, respectively, and spectra are box-car smoothed over 3
bins. The zero points of the spectra are shifted up for each phase,
but the scales are the same.  The exposure time is not the same for
each phase bin, so integrated counts will differ. For each profile,
the instrumental line profile (dash-dotted) is shown for comparison,
with peak set to the data peak and centroid set at the laboratory
wavelength.
}
\end{figure}

\begin{figure}
\plotone{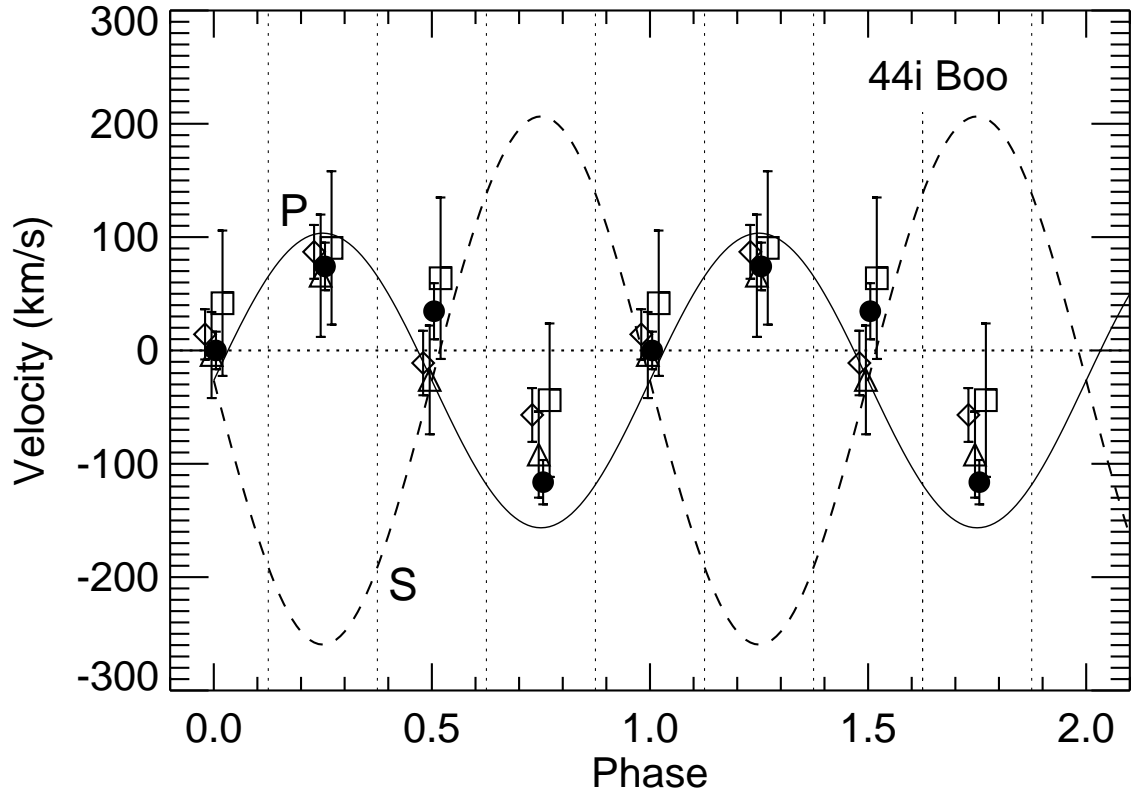}
\caption{Velocities as a function of orbital phase, as measured from
MEG profiles. Velocities are derived from the median centroid shifts
for the phase bins indicated by vertical dashed lines.  The four
symbols represent \ion{O}{8} ($\diamond$), \ion{Ne}{10} ($\bullet$),
\ion{Fe}{17} ($\Delta$), and \ion{Mg}{12} ($\square$). Error bars
represent statistical errors only. The different symbols are shifted
slightly from their phase centers for plotting.  Radial velocity
curves for the primary (solid) and secondary (dash-dotted) are from
Hill et al. (1989).
}
\end{figure}

\begin{figure}
\plotone{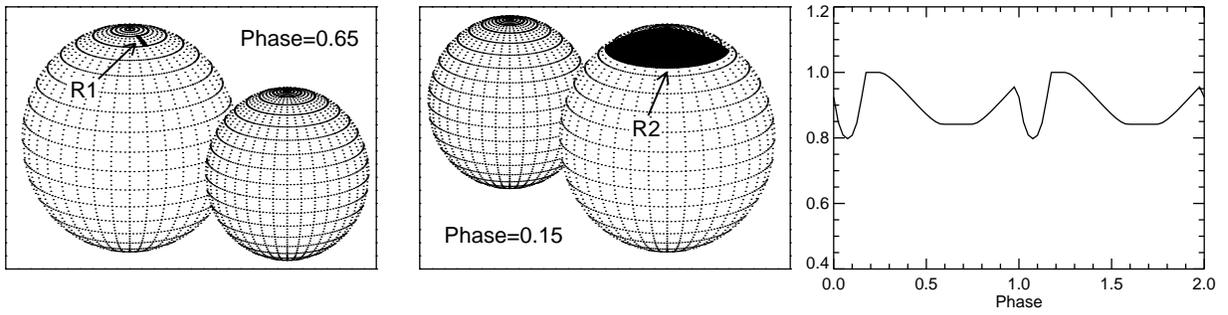}
\caption{Simulation of two active regions R1 and R2, described in
text, with model light curve.
  }
\end{figure}

\end{document}